\documentclass[aps,prl,twocolumn,showpacs,groupedaddress,floatfix,nofootinbib]{revtex4-1}
\usepackage{graphicx}
\usepackage{bm}				
\usepackage{amsfonts}
\usepackage{amssymb}
\usepackage{amsmath}
\usepackage{ulem} 
\usepackage{graphicx}		
\usepackage{xcolor, soul}	
\usepackage{verbatim}
\usepackage[T1]{fontenc}
\usepackage{pifont}
\usepackage{multirow}
\usepackage{notes2bib}
\usepackage{bibunits}
\usepackage{appendix}

\usepackage{MnSymbol}
\usepackage{makecell}
\usepackage{dsfont}
\usepackage{url}
\usepackage{rotating}
\usepackage{wasysym}
\usepackage{diagbox}
\let\oldAA\AA
\renewcommand{\AA}{\text{\normalfont\oldAA}}

\usepackage[colorlinks=true,linkcolor=black,citecolor=blue,urlcolor=blue,hypertexnames=false]{hyperref}
\newcommand{\nocontentsline}[3]{}
\newcommand{\tocless}[2]{\bgroup\let\addcontentsline=\nocontentsline#1{#2}\egroup}

\def\ba#1\ea{\begin{align}#1\end{align}}
\def\bg#1\eg{\begin{gather}#1\end{gather}}
\def\bpm{\begin{pmatrix}}
\def\epm{\end{pmatrix}}

\newcommand{\hgf}{HgF$_2$}

\newcommand{\bb}[1]{{\boldsymbol #1}}

\newcommand{\ct}{$\tilde C_{2z}T$}

\newcommand{\ourtitle}{
Euler band topology in spin-orbit coupled magnetic systems
}

\allowdisplaybreaks

\begin{document}
    \title{\ourtitle}
	
    \author{Seung Hun Lee$^{1,2,3}$}
    \altaffiliation{These authors contributed equally to this work.}
    \author{Yuting Qian$^{1,2,3*}$}
    \email{yuting_qian@snu.ac.kr}
    \author{Bohm-Jung Yang$^{1,2,3}$}
    \email{bjyang@snu.ac.kr}
    \affiliation{$^1$Department of Physics and Astronomy, Seoul National University, Seoul 08826, Korea\\
    	$^2$Center for Theoretical Physics (CTP), Seoul National University, Seoul 08826, Korea\\
    	$^3$Institute of Applied Physics, Seoul Nation University, Seoul 08826, Korea}
	
    \date{\today}
	  
    \begin{abstract}
    The Euler class characterizes the topology of two real bands isolated from other bands in two-dimensions. 
    Despite various intriguing topological properties predicted up to now, the candidate real materials hosting electronic Euler bands are extremely rare.
    Here, we show that in a quantum spin Hall insulator with two-fold rotation $C_{2z}$ about the $z$-axis, a pair of bands with nontrivial $\mathbb{Z}_2$ invariant turn into magnetic Euler bands under in-plane Zeeman field or in-plane ferromagnetic ordering. 
    The resulting magnetic insulator generally carries a nontrivial second Stiefel-Whitney invariant.
    In particular, when the topmost pair of occupied bands carry a nonzero Euler number, the corresponding magnetic insulator can be called a magnetic Euler insulator.
    Moreover, the topological phase transition between a trivial magnetic insulator and a magnetic Stiefel-Whitney or Euler insulator is mediated by a stable topological semimetal phase in which Dirac nodes carrying non-Abelian topological charges exhibit braiding processes across the transition.
    Using the first-principles calculations, we propose various candidate materials hosting magnetic Euler bands. We especially show that ZrTe$_5$ bilayers under in-plane ferromagnetism are a candidate system for magnetic Stiefel-Whitney insulators in which the non-Abelian braiding-induced topological phase transitions can occur under pressure.
    \end{abstract}

    \maketitle

    \indent
    \textit{Introduction.}---
    A pair of bands in two-dimensional (2D) systems with space-time inversion $I_{ST}$ symmetry can carry an integer Euler number $e_2$, which is a representative topological invariant characterizing magnetic band topology protected by antiunitary symmetries~\cite{ahn2019failure}.
    $I_{ST}$ symmetry in 2D systems appears in the form of $I_{ST}=C_{2z}T$ in spin-orbit coupled systems, while both $I_{ST}=C_{2z}T$ and  $I_{ST}=PT$ are possible in spinless cases where $C_{2z}$, $P$, and $T$ indicate two-fold rotation about the $z$-axis, inversion, and time-reversal symmetries, respectively. 
    Since $I_{ST}$ symmetry is local in momentum space and satisfies $I_{ST}^2=+1$, in $I_{ST}$ symmetric systems, one can always choose a real gauge in which both the Hamiltonian and wave functions are real~\cite{fang2015new,zhao2017p,ahn2018band,ahn2019}. Thus, the $I_{ST}$ symmetric systems are ideal playgrounds to study the topological characteristics of real wave functions described by the Euler class and the Stiefel-Whitney classes~\cite{hatcher2003vector,ahn2019}.

    Explicitly, for real two bands $|\tilde{u}_{m\bm{k}}\rangle$ ($m=1,2$) isolated by band gaps from others, $e_2$ is given by~\cite{ahn2019failure}
    \begin{equation}
    	e_2=\frac{1}{2\pi}\oint_{\text{BZ}} d\bm{S}\cdot\tilde{\bm{F}}_{12}(\bm{k}),
    \end{equation}
    where $\tilde{\bm{F}}_{12}(\bm{k})=\nabla_{\bm{k}}\times\tilde{\bm{A}}_{12}$ is the Euler curvature, $\tilde{\bm{A}}_{12}(\bm{k})=\langle \tilde{u}_{1\bm{k}}|\nabla_{\bm{k}}|\tilde{u}_{2\bm{k}}\rangle$ is the Euler connection, and $\text{BZ}$ denotes the Brillouin zone.
    Two real bands with $e_2\neq0$, shortly \textit{Euler bands}, exhibit various intriguing topological properties. For example, Euler bands violate the conventional fermion doubling theorem in a way that two Euler bands with $e_2$ accompany multiple nodal points in between whose total vorticity (or winding number) is $2e_2$~\cite{ahn2019failure}. 
    Moreover, $e_2$ of Euler bands can be changed via a band inversion with other bands, which is mediated by the pair-creation and pair-annihilation of multiple nodal points in different band gaps accompanied by relative braiding processes~\cite{ahn2019failure}. Since one-dimensional (1D) $I_{ST}$-symmetric multi-band systems carry non-Abelian topological charges, the braiding process of nodal points in 2D systems can also be described in terms of the braiding of non-Abelian topological charges~\cite{wu2019non,bouhon2020non}.
    
  \begin{figure}[h!]
  	\centering
  	\includegraphics[width=1\linewidth]{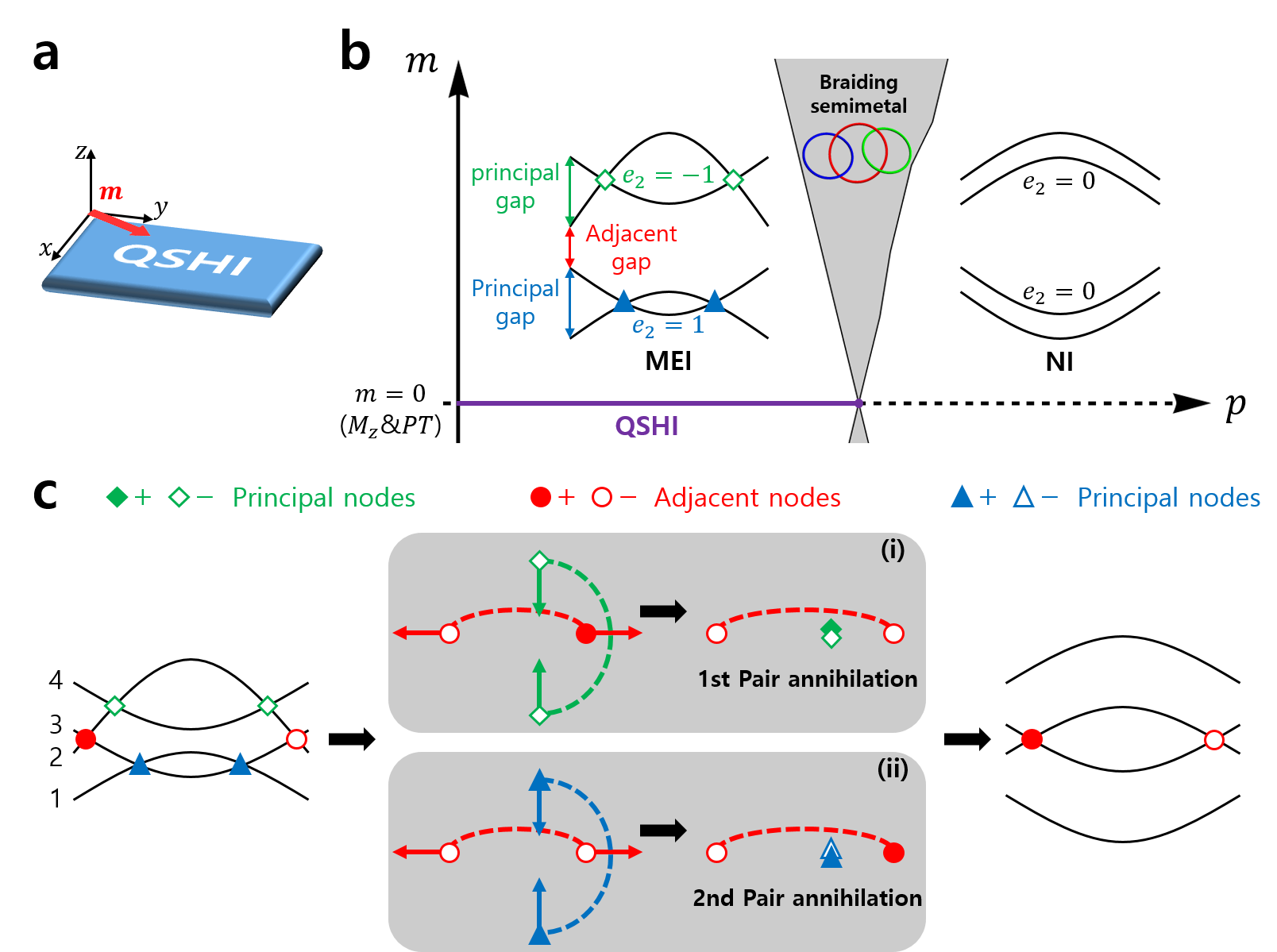}
  	\caption{
        {\bf {A mechanism to obtain magnetic Euler insulator (MEI), topological phase diagram, and related nodal braiding process.}}
  		\textbf{a} The schematic figure for a MEI obtained by introducing in-plane magnetism ($m$) to a quantum spin Hall insulator (QSHI).
  		\textbf{b} The topological phase diagram from the Bernevig-Hughes-Zhang (BHZ) model with in-plane magnetism, 
  		and the typical band structure of MEI and normal insulator (NI).
  		The filled (empty) blue, red, and green markers represent the nodes between the band 1 and 2, 2 and 3, 3 and 4 with vorticity $+1$ ($-1$), respectively. 
  		\textbf{c} A typical band structure of the braiding semimetal and a schematic description of nodal braiding processes.
  		The dashed lines represent the Dirac strings that connect two nodes with the same color.
  	}\label{fig1}
  \end{figure}
      
    After the first theoretical discovery of the Euler bands in magic-angle twisted bilayer graphene~\cite{ahn2019failure,po2019faithful,song2019all}, the Euler band topology and the related momentum space braiding processes have been theoretically predicted in 
    electronic bands of three-dimensional (3D) Weyl semimetals or nodal line semimetals such as scandium, ZrTe, and Cd$_2$Re$_2$O$_7$~\cite{wu2019non,bouhon2020non,chen2022non}, and further investigated in acoustic phonons~\cite{park2021topological} and phonon bands in layered materials Si$_2$O$_3$ and Al$_2$O$_3$~\cite{peng2022phonons,peng2022multigap}.
    Recently, significant advances in the study of the Euler band topology and nodal point braiding process were made in artificial lattice systems including optical lattices \cite{Slager20prl}, acoustic metamaterials~\cite{peri2020experimental,jiang2021experimental,jiang2022experimental,qiu2023minimal,davoyan2023mathcal}, photonic metamaterials or crystals \cite{park2021non,photonicmata}, transmission line networks \cite{guo2021exp,jiang2021four}, etc. 
    However, the lack of 2D solid materials hosting electronic Euler bands under spin-orbit coupling has limited the exploration of the Euler band topology in condensed matter systems.
    
    In this letter, we introduce a simple and general route to obtain the Euler bands in 2D spin-orbit coupled electronic systems: the application of in-plane magnetism to the $C_{2z}$-symmetric quantum spin Hall insulator (QSHI) as schematically described in Fig.~\ref{fig1} (a).
    Here, the in-plane magnetism includes external Zeeman magnetic fields, intrinsic magnetic orderings, and also extrinsic magnetism from proximate magnetic heterostructures. 
    We dub the resulting $T$-broken insulator as a magnetic Euler insulator (MEI) if it hosts isolated Euler bands as the topmost occupied bands (See Fig.~\ref{fig1} (b)). 
    Even if the Euler bands are not decoupled from other bands, since the occupied bands still carry a nonzero second Stiefel-Whitney (SW) number $w_2$, the corresponding insulator can be called a 2D magnetic SW insulator (MSWI).
    In general, isolated two bands with nontrivial Kane-Mele $\mathbb{Z}_2$ invariant $\nu_2$ in $C_{2z}$ and $T$-symmetric 2D systems turn into magnetic Euler bands under the in-plane magnetism.
    Using a lattice model, we demonstrate that the topological phase transition between the MEI/MSWI and a normal insulator (NI) is mediated by a semimetal phase, dubbed a braiding semimetal, in which the braiding of nodal points in different band gaps occurs across the transition (See Fig.~\ref{fig1} (c)). 
    Based on first-principles calculations, we report that such nodal point braiding in 2D electronic bands can be realized in a MSWI candidate, the bilayer ZrTe$_5$ with in-plane ferromagnetism under pressure.

\indent
\textit{Band topology.}---
To deliver the central idea, let us briefly recap the relation among three distinct topological invariants: the Euler number $e_2$, the second SW number $w_2$, and the Kane-Mele invariant $\nu_2$.
$\nu_2$ is a $\mathbb{Z}_2$ topological invariant for 2D spinful $T$-symmetric systems distinguishing a QSHI ($\nu_2=1$) and a NI ($\nu_2=0$)~\cite{kane2005quantum,bernevig2006quantum}.
On the other hand, $w_2$ is a $\mathbb{Z}_2$ topological invariant for $I_{ST}$-symmetric real $N$-band systems ($N\geq2$), equivalent to the parity of $e_2$ when $N=2$~\cite{ahn2019failure,ahn2018band,ahn2019,hatcher2003vector}.
Mathematically, $w_2$ describes whether a spin or pin structure is allowed or not for given real wave functions defined on a 2D closed manifold~\cite{ahn2018band,ahn2019,hatcher2003vector}. 
Physically, it describes a $\mathbb{Z}_2$ monopole charge of a nodal line in spinless $PT$-symmetric systems~\cite{ahn2018band,fang2015new,zhao2017p}, or a $\mathbb{Z}_2$ topological invariant distinguishing a 2D SW insulator ($w_2$=1) and a NI ($w_2=0$) ~\cite{ahn2018band,ahn2019}.
(See the Supplemental Materials (SM) S1~\cite{supplement} for more details).
	
When $T$-symmetry exists, $w_2$ is also equivalent to $\nu_2$~\cite{ahn2019failure}.
Therefore, a $C_{2z}$-symmetric QSHI with two occupied bands can be regarded as an Euler insulator.
If $T$-symmetry is broken while $C_{2z}T$-symmetry is preserved, the QSHI turns into a MEI as long as $T$-breaking does not close the bulk gap. 
Therefore, the MEI can be realized by introducing in-plane magnetism into the $C_{2z}$-symmetric QSHI.

\indent
\textit{Model.}---
To illustrate the idea, we employ the Bernevig-Hughes-Zhang (BHZ) model for QSHI~\cite{bernevig2006quantum,konig2007quantum}
\begin{align}
	H_{\textrm{BHZ}}(\bm{k})=
	\begin{pmatrix}
		\varepsilon_s-2t_{ss}f(\bm{k})	& 2t_{sp}g(\bm{k})\\
		2t_{sp}g^*(\bm{k})	& \varepsilon_p-2t_{pp}f(\bm{k})
	\end{pmatrix},
\end{align}
where $\varepsilon_s$ and $\varepsilon_p$ denote the on-site potential of the $s$- and $p$-like orbitals, $t_{ss}$ and $t_{pp}$ denote the intra-orbital hoppings, and $t_{sp}$ denotes the inter-orbital hopping, respectively~\cite{bernevig2006quantum}.
Here, $f(\bm{k})=(p_0-p)(\cos{k_x}+\cos{k_y})$, $g(\bm{k})=(p_0-p)(\sin{k_y}\sigma_z-i\sin{k_x}\sigma_0)$, and $p_0=\left|\varepsilon_s-\varepsilon_p\right|/4(t_{ss}+t_{pp})$.
$\sigma_{x,y,z}$ and $\sigma_0$ are the spin Pauli matrices and the corresponding identity matrix, respectively.
We note that $H_{\textrm{BHZ}}(\bm{k})$ describes a QSHI (NI) when $p>0$ ($p<0$).
$p$ is a parameter, such as pressure, that modulates the hoppings' strength and drives a topological phase transition (TPT).

Now we take the in-plane magnetism into account and obtain $H_{\textrm{MBHZ}}(\bm{k})\equiv H_{\textrm{BHZ}}(\bm{k})-\bm{m}\cdot\bm{\sigma}$, where $\bm{m}=(m_x,m_y,0)$ indicates the effective magnetic moment.
The resulting phase diagram as a function of $p$ is shown in Fig.~\ref{fig1} (b).
$|\bm{m}|\equiv m\neq0$ turns the QSHI at $m=0$ into a MEI as long as $m$ does not close the bulk gap.
Also, as shown in Fig.~\ref{fig1} (b), the lower (upper) two bands of the MEI with $e_2=+1$ ($e_2=-1$) possess two Dirac points with the same vorticity $+1$ ($-1$) while the NI with $e_2=0$ has no node, satisfying the generalized Nielsen-Ninomiya theorem in the form of $N_{v}^{\text{total}}=2e_2$ where $N_{v}^{\text{total}}$ indicates the total sum of the vorticities of the Dirac points between the two bands with $e_2$~\cite{ahn2019failure}.

\begin{figure*}[t!]
	\centering
	\includegraphics[width=0.98\textwidth]{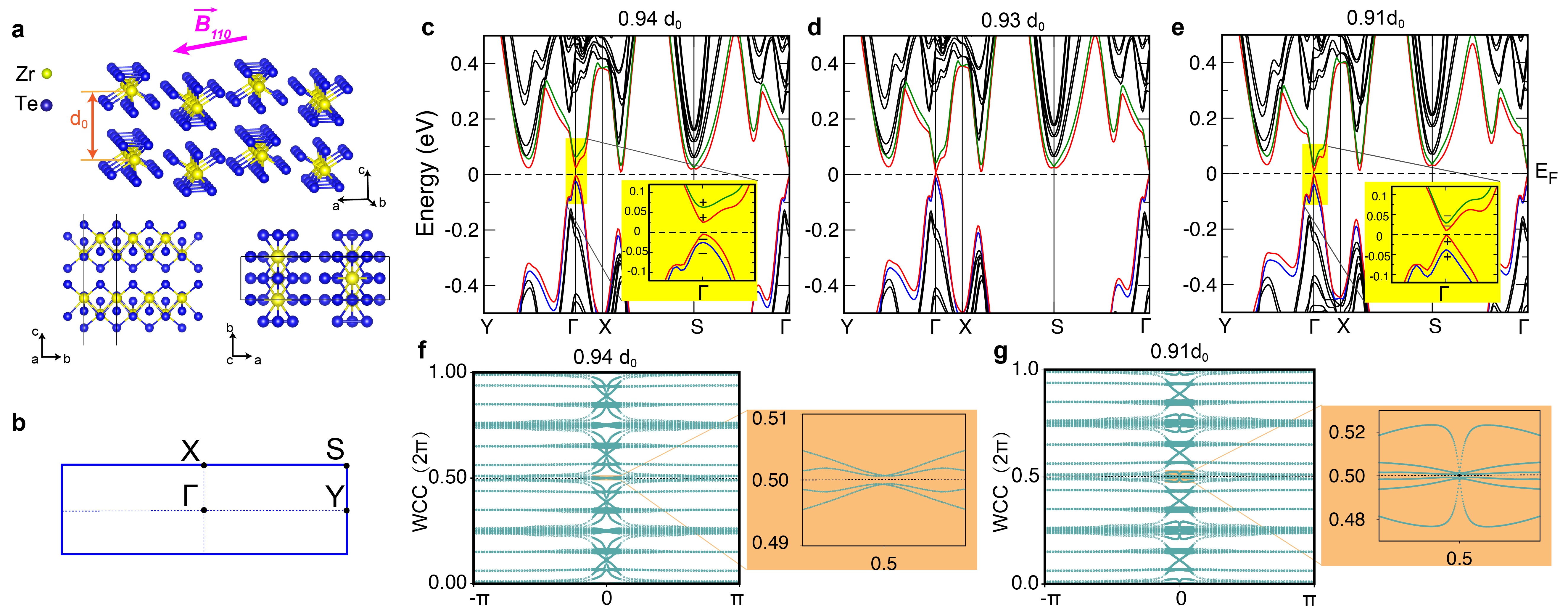}
	\caption{
		{\bf Band topology of bilayer ZrTe$_5$ with in-plane magnetism.}
		\textbf{a} Crystal structure of bilayer ZrTe$_5$ considering in-plane magnetism along (110)-direction.
		The top panel shows the side view from the $b$ axis and the original layer distance labeled by $ d_0$.
		The bottom panel shows the side view from the $a$ axis and the top view with the primitive cell represented by black lines.
		\textbf{b} 2D Brillouin zone and high symmetry $\bf k$-points.
		\textbf{c-e} The spin-orbit coupled band structures show a TPT when the layer distance changes as $\lambda=0.94  d_0$, $0.93  d_0$, and $0.91  d_0$.
		A gapless phase appears at the critical point with $\lambda=0.93  d_0$, while gapped phases appear for $\lambda=0.94  d_0$ and $0.91  d_0$.
		The inset of \textbf{c,e} shows the four bands around $\Gamma$ near the Fermi level labeled by inversion eigenvalues $\pm1$.
		\textbf{f} Wilsonloop spectra on 2D Brillouin zone for $\lambda=0.94  d_0$ gapped phase and a close-up of orange regions in the right panel, indicating $w_2=0$.
		\textbf{g} Similar plot for $\lambda=0.91  d_0$ gapped phase exhibiting $w_2=1$.
	}\label{fig:zrte}
\end{figure*}
	
    \indent
    \textit{Topological phase transition (TPT) and braiding of nodes.}---
     Let us describe the TPT.
     When $m=0$ with $M_z:z\rightarrow -z$ mirror symmetry, each band is doubly degenerate due to $P=C_{2z}M_{z}$ and $T$ symmetries, and the TPT occurs directly through a single critical point~\cite{murakami2007phase,murakami2007tuning,murakami2008}. 
     On the other hand, when $m\neq0$, the TPT between the MEI and NI goes through a stable semimetal phase. 
     In particular, the gap-closing node in the 2D momentum $(k_x,k_y)$ space at $m=0$ deforms to three nodal rings in the 3D $(k_x,k_y,p)$ space corresponding to the trajectory of gap-closing nodes between the bottom two bands (band 1, 2), two middle bands (band 2, 3), and top two bands (band 3, 4), respectively, as $p$ is varied (See Fig.~\ref{fig1} (b)).
     The linking structure of the middle ring with the other two rings describes the braiding process of nodes in the 2D momentum space (See SM S3)~\cite{supplement}.
    
    Generally, a TPT changing $e_2$ is involved with the braiding of nodes in different band gaps~\cite{ahn2019failure,po2019faithful,bouhon2022multi}.
    Since the occupied bands have nodes with the same vorticity in MEI while they have no nodes in NI, the semimetal phase should mediate the node annihilation in the occupied bands, which occurs through the sign reversal of the vorticity in one of the two nodes by crossing Dirac strings (See SM S2)~\cite{supplement}. 
    
    Let us elaborate on the idea as follows. 
    In general, an $I_{ST}$-symmetric two-band Hamiltonian can be written as
    \begin{align}
    	H_{2\times2}(\bm{k})=r(\bm{k})\cos\theta(\bm{k})\sigma_x+r(\bm{k})\sin\theta(\bm{k})\sigma_z,
    \end{align} 
    where $r(\bm{k})\geq0$ and $\theta(\bm{k})\in[0,2\pi]$ are real numbers, $\sigma_{x,z}$ are Pauli matrices defined in the real two-band basis assuming the representation $I_{ST}=K$ with the complex conjugation operator $K$. The unimportant term proportional to the identity matrix is neglected.
    The vorticity $N_{v}$ of a Dirac point is nothing but the winding number of the Hamiltonian $H_{2\times2}(\bm{k})$ along a loop $C$ encircling the Dirac point, which is explicitly given by $N_v=\frac{1}{2\pi}\oint_{C}d\bm{k}\cdot\nabla_{\bm{k}}\theta(\bm{k})$.

    In Fig.~\ref{fig1} (b), the pair of blue (green) nodes between the band 1 and 2 (3 and 4) in the MEI have the same vorticity.
    As $p$ increases, a crossing between the band 2 and 3 creates a pair of red nodes with opposite vorticities.
    As $p$ increases further, one of the blue (green) nodes encircles one red node and experiences the sign reversal of its vorticity as described in the middle panel of Fig.~\ref{fig1} (c).
   The vorticity reversal happens because, for the blue (green) nodes, a red node is a source of $\pi$ Berry phase when encircled. Because of this, the real wave function for the occupied (unoccupied) bands develops a singularity at a point on any closed path enclosing a red node~\cite{ahn2019failure}. The Dirac string is nothing but a line of such singularities between two red nodes, which appears in the form of the branch cut for $\tilde{\bm{A}}(\bm{k})$~\cite{ahn2019failure,ahn2018band}.
   In general, for a node (principal node), a node in the adjacent gap (adjacent node) is a source of $\pi$ Berry phase, thus when a principal node encircles an adjacent node, the principal node always crosses a Dirac string between two adjacent nodes and experiences its vorticity sign reversal.
   Since the pair of blue (green) nodes have opposite vorticities after the vorticity reversal, their pair annihilation becomes possible.
   We note that during the whole braiding process, one of the red nodes passes through two Dirac strings (one between two blue nodes and the other between two green nodes), recovering its original vorticity.
   After the pair-annihilation of blue (green) nodes, with further increasing $p$, the red nodes also undergo pair-annihilation, leading to the NI with $e_2=0$.

\indent
\textit{First-principles calculation.}---
There are several candidate materials for $C_{2z}$-symmetric QSHI such as  HgTe/CdTe quantum wells~\cite{bernevig2006quantum,konig2007quantum}, transition metal compounds ZrTe$_5$~\cite{weng2014} and Ta$_2$$M_3$Te$_5$ ($M$ = Ni, Pd)~\cite{guo2021prb}, which can host magnetic Euler band topology under in-plane magnetization.  Additionally, a diamond-octagon flat band family $M$F$_2$ ($M$ = Zn, Cd, and Hg)~\cite{2dmatpedia, HgF2prm}, although not categorized as a QSHI, features two isolated QSH bands near the Fermi energy, making it conducive for studying magnetic Euler topology as shown in SM S7~\cite{supplement}. 
To study the braiding process, we have found that bilayer ZrTe$_5$ is the best system because the band topology is tunable by controlling the inter-layer distance.
ZrTe$_5$ has recently gained attention as a topological material in both theoretical and experimental studies~\cite{weng2014,tang2019}. First-principles calculations have shown that 3D bulk ZrTe$_5$ is close to the boundary between weak and strong topological insulators, making it an ideal platform for investigating topological phase transitions~\cite{weng2014}.

\begin{figure*}[t!]
	\centering
	\includegraphics[width=0.98\textwidth]{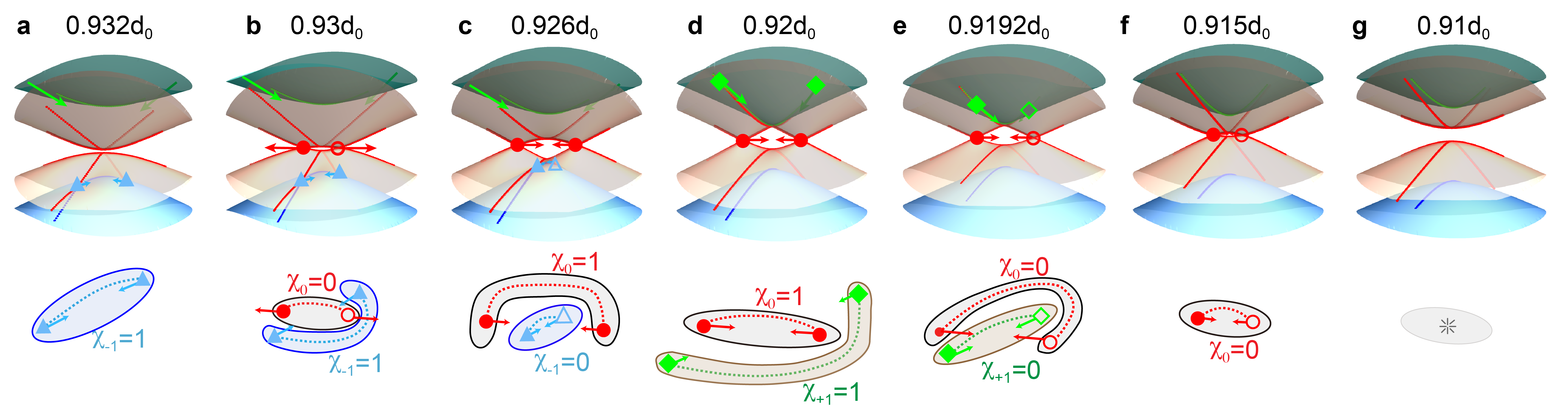}
	\caption{
		{\bf Braiding of nodes in bilayer ZrTe$_5$ with in-plane ferromagnetism by varying the layer distance.}
		\textbf{a-g} Braiding between principal nodes (green, blue) and adjacent nodes (red) around $\Gamma$ as $\lambda$ varies, obtained from first-principles calculations. The upper panel is the band structure around $\Gamma$ with $k_1,k_2 \in [-0.03 \frac{2\pi}{a}:0.03 \frac{2\pi}{b}]$ where $a$ and $b$ indicates the in-plane lattice constants. The lower panel shows the top view of the braiding of nodes, along with their charges and related patch Euler classes.
		The dashed lines connecting two nodes represent the Dirac strings.
		The arrows indicate the moving direction of nodes.
	}
	\label{fig:braid_mat}
\end{figure*}

The crystal structure of ZrTe$_5$ adopts a layered configuration with the $Cmcm$ space group (No. 63)~\cite{zrte_str}. Interestingly, even a single layer of ZrTe$_5$ exhibits exceptional properties as a QSHI, featuring a substantial energy gap of approximately 100 meV~\cite{weng2014,wu2016evidence,li2016exp}. The robustness of the QSHI state against strain is attributed to the fact that the nontrivial band inversion can only be trivialized by strong inter-chain coupling at the $\Gamma$ point between the $p$-orbitals of Te dimer atoms~\cite{weng2014}.
Notably, due to the extremely weak van der Waals interlayer coupling~\cite{weng2014}, it becomes feasible to manipulate the inter-chain coupling of the $p$-orbital of Te dimer atoms by adjusting the inter-layer distance experimentally, thus enabling the realization of topological phase transitions in two-dimensions.
Furthermore, recent transport experiments have detected Zeeman splitting in ZrTe$_5$ under magnetic fields~\cite{liu2016zeeman}. Given these remarkable characteristics of ZrTe$_5$, one can expect that the braiding of nodes can be achieved in 2D bilayer ZrTe$_5$ under the influence of an external in-plane Zeeman field.

In Fig.~\ref{fig:zrte} (a) and (b), the crystal structure and BZ of a 2D bilayer ZrTe$_5$ with the original layer distance ${d_0}=7.265 \AA$~\cite{zrte_str} are depicted.
Without in-plane magnetism, the electronic structure of the 2D bilayer ZrTe$_5$ is topologically trivial. 
However, upon reducing the layer distance $\lambda$, a TPT to a QSHI is observed (See SM S6)~\cite{supplement}.
Subsequently, a weak in-plane Zeeman field is introduced along the (110) direction.
In Fig.~\ref{fig:zrte} (c-e), it is evident that a phase transition occurs as the layer distance decreases from $\lambda=0.94d_0$ to $0.91d_0$, where the gap between the highest occupied and the lowest unoccupied band (highlighted in red) closes at $\lambda\approx0.93d_0$.
To confirm the TPT, we compute the topological invariants before and after the band inversion. 
%
First, we have confirmed that the Berry phase vanishes, so $e_2$ and $w_2$ can be well-defined.
Since we deal with multi-band systems, we compute $w_2$ by using the Wilson loop method, as shown in Fig.~\ref{fig:zrte} (f) and (g). For $\lambda = 0.94d_0$, the absence of Wilson band crossings at $\theta=\pi$ gives $w_2=0$~\cite{ahn2018band}. Meanwhile, for $\lambda=0.91d_0$, the Wilson band crossing at $\theta=\pi$ indicates that this system is a MSWI with $w_2=1$. This phase transition is accompanied by a double band inversion, i.e., the inversion between two occupied and two unoccupied bands, which is also confirmed by counting the inversion eigenvalues at time-reversal invariant momenta~\cite{ahn2018band}, as shown in the insets of Fig.~\ref{fig:zrte} (c) and (e).

Now, we show that the TPT is mediated by the braiding processes between the principal and adjacent nodes. 
To gain a deeper understanding of the braiding process, it is essential to meticulously track the trajectories of the principal nodes 
and the adjacent nodes 
as we vary the layer distance $\lambda$.
As shown in Fig.~\ref{fig:braid_mat}, gapless phases manifest within the range of $\lambda \in [0.93,0.915] d_0$.
Remarkably, two distinct braiding processes become apparent as we decrease $\lambda$ from $0.93d_0$ to $0.915d_0$. To trace the evolution of node vorticities, we employ the calculation of the patch Euler class~\cite{bouhon2020non,peng2022phonons}, the Euler class integral calculated inside a small region (patch) around the nodes of interest (See SM S1)~\cite{supplement}.
In the ensuing discussion, we assign patch Euler numbers based on three energy gaps near the Fermi level, denoting them as {$\chi_{-1}$, $\chi_0$, $\chi_1$} in ascending order of energy.
At $\lambda=0.932d_0$, an adjacent gap remains, hosting two pairs of principal nodes below and above the Fermi level near the $\Gamma$ point. In this case, the patch Euler numbers $\chi_{-1}=\chi_1=1$ reveal that each pair of green and blue nodes in the principal gaps carry the same charge, as illustrated in Fig.~\ref{fig:braid_mat}~(a). 
(Note that the green nodes away from $\Gamma$ are outside the area we plot in Fig.~\ref{fig:braid_mat} (a)).
When $\lambda=0.93d_0$, two adjacent nodes emerge in the gap between two red bands with a patch Euler number $\chi_0=0$, signifying that these newly created nodes possess opposite vorticities.
Subsequently, as the layer distance is further reduced, the two red nodes move away from the $\Gamma$ point, while the two lower principal blue nodes simultaneously approach the $\Gamma$ point. At $\lambda=0.926d_0$, one red node and one blue node flip the signs of their vorticities after crossing each other's Dirac string, changing the corresponding patch Euler classes as well. 
At this stage, two red nodes cannot be annihilated within the grey patch. However, a pair of principal blue nodes can be further annihilated, as observed at $\lambda=0.92d_0$.
As the layer distance decreases further, the two red nodes move towards the $\Gamma$ point, while the two green nodes also approach the $\Gamma$ point faster. Ultimately, the second braiding process takes place as $\lambda$ reaches $0.9192 d_0$, causing the nodes in two different gaps to reverse the sign of their vorticities, as depicted in Fig.~\ref{fig:braid_mat}~(e).
Consequently, as we further decrease $\lambda$, two principal green nodes are annihilated at $\lambda=0.915d_0$, and two adjacent red nodes are also annihilated at $\lambda=0.91d_0$.

We note that the four bands of interest in ZrTe$_5$ are not completely separated from other bands. Nevertheless, there are significant gaps between the four bands and the other bands in the vicinity of $\Gamma$ where the braiding processes take place. 
Therefore, the braiding processes of nodes mediating the TPT are clearly visible. 
We propose that a similar process can also be studied in Ta$_2$$M_3$Te$_5$ family with pressure under in-plane magnetic field~\cite{guo2022}.

\indent
\textit{Discussion.}---
To conclude, we have demonstrated that the in-plane magnetism turns a $C_{2z}$-symmetric QSHI into a spinful $T$-broken MEI, and the TPT between a MEI and NI is mediated by a semimetal phase in which various braiding processes of nodes appear.  
We note that the in-plane magnetism does not have to be ferromagnetic as long as it preserves $C_{2z}T$.
For instance, the MEI and related nodal braiding can also be observed in non-collinear antiferromagnets on the spin-orbit coupled kagome lattice.
As shown in SM, for both the ferromagnetic and antiferromagnetic cases, the isolated two bands with non-trivial $e_2$ and related nodal braiding processes mediating TPTs appear in the kagome lattice. 
We expect that our proposal for electronic Euler bands in spin-orbit coupled magnetic systems will stimulate the future study of real band topology and related momentum space braiding in condensed matter systems.

\begin{acknowledgments}
We thank Sunje Kim, Soonhyun Kwon, Rasoul Ghadimi, Yan Zhang and Quansheng Wu for the fruitful discussions. S.L., Y. Q., B.-J.Y. were supported by Samsung  Science and Technology Foundation under Project Number SSTF-BA2002-06, the National Research Foundation of Korea (NRF) grant funded by the Korean government (MSIT) (No.2021R1A2C4002773, and No. NRF-2021R1A5A1032996).
\end{acknowledgments}

\bibliographystyle{apsrev4-1}
\bibliography{Refs}


\clearpage

\onecolumngrid

\appendix


\renewcommand{\appendixpagename}{\center\large Supplemental Material for ``\ourtitle''}

\appendixpage

\setcounter{page}{1}
\setcounter{section}{0}
\setcounter{figure}{0}
\setcounter{equation}{0}
\setcounter{table}{0}

\renewcommand{\thefigure}{S\arabic{figure}}
\renewcommand{\theequation}{S\arabic{equation}}
\renewcommand{\thesection}{S\arabic{section}}
\renewcommand{\thetable}{S\arabic{table}}

\noindent
\indent In this Supplemental Material, we provide detailed discussions on the results presented in the main text. The sections are organized as follows.\\\\
\textbf{S1. Method} \hfill \pageref{method}\\\\
\textbf{S2. General effective four-band Hamiltonian describing the gap closing} \hfill \pageref{gapclosing}\\\\
\textbf{S3. Nodal braiding in the BHZ model} \hfill \pageref{BHZ}\\\\
\textbf{S4. Kagome lattice ferromagnet} \hfill \pageref{kagomeFM}\\\\
\textbf{S5. Kagome lattice non-collinear antiferromagnet} \hfill \pageref{kagomeAFM}\\\\
\textbf{S6. Topology of ZrTe$_5$ systems} \hfill \pageref{app:zrteband}\\\\
\textbf{S7. Magnetic Euler topology in HgF$_2$ systems} \hfill \pageref{app:hgf2band}\\\\

\section{S1. Method}\label{method}
{\bf Computation details.}
The first-principles calculations were performed with the Vienna Ab initio Simulation Package (VASP) package~\cite{KRESSE199615,vasp} based on the density functional theory (DFT). 
We used the crystal structure of ZrTe$_5$ obtained from the experiment~\cite{zrte_str}. 
To check the stability of the crystal structure, we have performed a relaxation until the force on each atom is less than 0.01 eV/$\AA$. 
No relaxation is considered when analyzing the braiding process by tuning the layer distance.
Crystal structures were plotted by VESTA package~\cite{Vesta}.
The generalized gradient approximation (GGA) of the Perdew-Burke-Ernzerhof type is employed for exchange-correlation potential~\cite{pbe96,perdew1998perdew}. 
The kinetic energy cutoff was set to 500 eV.
For Brillouin zone (BZ) sampling in the self-consistent calculation, an $8 \times 8 \times 2$ $k$-point mesh was adopted for 3D bulk ZrTe$_5$, and $3 \times 10 \times 1$ $k$-point for 2D single layer ZrTe$_5$, and $2 \times 8 \times 1$ $k$-point for 2D bilayer ZrTe$_5$. 

To simulate the Zeeman effect and tune the layer distance, we construct two Wannier tight-binding models with 120 orbitals for systems $H_{0.9}$ and $H_{1.0}$ with layer distance $\lambda=0.9d_0$ and  $\lambda=d_0$ based on Te $p$-orbitals by Wannier90 package~\cite{mlwf}. 
The intermediate systems with layer distance $\lambda \in (0.9,1.0)$ $d_0$ are obtained by interpolating the above two Hamiltonian as $H_{soc}=m H_{1.0} + (1-m)H_{0.9}$, where $m=10(\lambda-0.9d_0)/d_0$.
After applying an external magnetic field, the total Hamiltonian is obtained by adding the Zeeman term $H_{zm}=\gamma \vec{\bb B} \cdot \vec{\bb \sigma} \equiv \gamma (B_x\sigma_x \cos(\phi) +B_y\sigma_y\sin{\phi})$ into $H_{soc}$, ie., $H=H_{soc}+H_{zm}\cdot I_{60}$, where $I_{60}$ is an identity matrix with dimension $60\times 60$.
In our case, we apply a (110) magnetic field, i.e., fixing $\gamma = 0.02$ and $\phi=45^\circ$.
WannierTools~\cite{wu2018wanniertools} was further employed to calculate the Wilson-loop spectrum and find nodes within gaps.

{\bf Stiefel-Whitney class.}
The Stiefel-Whitney (SW) classes are characteristic classes of a real vector bundle that describe the obstructions to constructing everywhere independent sets of sections of the vector bundle.
The following information on the first and second SW classes is extracted from Ref.~\cite{ahn2019}.

The first SW class $w_1$ measures the orientability of a real vector bundle over a 1D closed manifold. If $w_1 = 0$ ($w_1 = 1$), the real vector bundle is orientable (non-orientable).
In general, the orientation of a real vector bundle refers to the choice of an ordered basis. For any two ordered bases, there exists a unique non-singular linear transformation that relates them. When the determinant of the transformation matrix is positive (negative), we say that the bases have the same (different) orientation. After choosing an ordered reference basis $\{v_1,v_2,...\}$, the orientation of another basis $\{u_1,u_2,...\}$ is specified to be positive (negative) when the basis has the same (different) orientation with the reference basis. Real wave functions defined on the BZ can be considered as real unit basis vectors defined at each momentum. Hence, they form a structure of a real vector bundle over the BZ. The basis can be smoothly defined locally on the manifold, but may not be smooth over a closed submanifold $\mathcal{M}$ of our interest. We say that the real wave functions are orientable over $\mathcal{M}$ when the local bases can be glued with transition functions with only positive determinants, i.e., all transition functions are orientation-preserving. The orientable wave functions are classified into two classes with positive and negative orientations as in the case of the real vector spaces. Interestingly, the orientability of real wave functions can be determined by the Berry phase computed in a smooth complex gauge, such that $w_1 = 1$ ($w_1 = 0$) indicates that the relevant wave functions carry $\pi$ ($0$) Berry phase.

The second SW class $w_2$ is a $\mathbb{Z}_2$ invariant, describing  whether a spin (or pin) structure is allowed for a real vector bundle on a 2D closed manifold $\mathcal{M}$. If $w_2 = 0$ ($w_2 = 1$), a spin or pin structure is allowed (forbidden).
Let $A$ and $B$ be two open covers of $\mathcal{M}$ whose geometric structure is topologically equivalent to $\mathcal{M}$. In $A$ and $B$, one can find smooth real wave functions $|u_{m\bm k}^A\rangle$ and $|u_{m\bm k}^B\rangle$ defined in each open cover, respectively ($m,n=1,2,\cdots,N$). In the overlapping region $A\cap B$, a transition function $t_{mn}^{AB}(\bm k)$ is defined as $|u_{n\bm k}^B(\bm k)\rangle=t_{mn}^{AB}(\bm k)|u_{m\bm k}^A\rangle$.
For convenience, we assume that the $N$ wave functions are orientable over $\mathcal{M}$ so $t_{mn}^{AB}(\bm k)\in\textrm{SO}(N)$. Since the double covering of $\textrm{SO}(N)$ is $\textrm{Spin}(N)$, the problem reduces to whether the lifting of the transition function
from $t\in \textrm{SO}(N)$ to $\tilde{t}\in\textrm{Spin}(N)$ is allowed or not. If such lifting is allowed (not allowed) consistently over $\mathcal{M}$, we say that the
spin structure exists (does not exist) and thus $w_2 = 0$ ($w_2 = 1$).
$w_2$ can be determined by counting the number of crossing points in the Wilson loop spectrum on the $\Theta = \pi$ line.
Physically, the second SW $w_2$ characterizes the $\mathbb{Z}_2$ monopole nodal charge in 3D nodal line semimetals and $\mathbb{Z}_2$ topological invariant of 2D insulators in the absence of the Berry phase.

{\bf Euler class.} The Euler class is a characteristic class of an oriented real vector bundle.
As introduced in Eq.~(1) of main text, the Euler invariant $e_2$ is a topological $Z$ invariant for two real bands. $e_2$ is invariant under orientation-preserving gauge transformations whose determinant is +1, while it reverses its sign under orientation reversing gauge transformations whose determinant is -1. Thus, $e_2$ is well-defined only for orientable states.
The band topology associated with the nonzero $e_2$ is fragile. Namely, the Wannier obstruction of two Euler bands with $e_2 \neq 0$ disappears when additional trivial bands are introduced. Although $e_2$ is defined only for two band systems, its parity still remains meaningful even after additional trivial bands are introduced. That is, if the Euler class of the two-band model is even (odd), the second SW $w_2$ of the system should remain zero (one) after the inclusion of additional trivial bands.

{\bf Non-Abelian charges.}
In materials research, topological classification has been extensively studied based on a single gap in the past decades. 
However, very recently, it has been found that when considering multiple gaps, there are some physical properties overlooked in single-gap studies~\cite{ahn2019failure,wu2019non,bouhon2020non}.
In such multi-gap real band systems, non-Abelian frame charges characterize their topology.
Essentially, these charges delineate the behavior of band nodes in systems with multiple bands.

For simplicity, let us consider a system comprising three non-degenerate real bands, where two gaps within the bands are defined.
In a real basis, Bloch eigenstates at a momentum $\bm k$ form an oriented orthogonal matrix that represents a \textit{frame}, i.e., $\mathcal{F}(\bm k)\equiv(|u_{1\bm k}\rangle, |u_{2\bm k}\rangle, |u_{3\bm k}\rangle)$. 
In general, such a frame can be considered to be rotated from the reference frame $\mathcal{F}_{\textrm{ref}}=\textrm{diag}(1,1,1)$ by a frame rotation $R \in \textrm{SO}(3)$ whose form reads $R=\exp[\alpha L_1+\beta L_2+\gamma L_3]$,
where $L_{i=x,y,z}$ are the rotation generators for three axes in the reference frame.
Let us pick two of these three bands and consider nodes in the gap between the two bands.
Then the frame acquires the total rotation angle $\phi=\sqrt{\alpha^2+\beta^2+\gamma^2}=\pi$ as we vary $\bm k$ along a closed path enclosing a single node.
Meanwhile, the accumulated frame charge on a loop enclosing a pair of nodes is 0 ($2\pi$) if the nodes can (cannot) be annihilated. Given that the $\pi$-rotations around the three axes anticommute with each other, the structure of frame charges is encoded in the quaternion group $\mathbb{Q}$ since for the real three bands, the space of spectrally flattened Hamiltonian on a loop that avoids nodes is given by O(3)$/\mathbb{Z}_2^3$ and $\pi_1[\textrm{O}(3)/\mathbb{Z}_2^3]=\mathbb{Q}$~\cite{wu2019non,bouhon2020non,peng2022phonons}. In other words, the topological charge of the nodes inside the loop becomes an element of $\mathbb{Q}$.

The band nodes thus carry frame charges $q \in \{1, \pm i, \pm j, \pm k, -1\}$, satisfying $(\pm 1)^2=1$, $ij=k$, $jk=i$, $ki=j$, and $i^2=j^2=k^2=ijk=-1$.

Let us consider the bands shown in Fig.~1 (c) as an example. We can assign a charge $\pm i$ to every node of the first energy gap represented by triangle 
and $\pm j$ to every node of the second energy gap represented by circle.
If one node with $q=i$ in the first gap rotates around a node with $q=j$ in the second gap, the charge of this node in the first gap will be changed to $(-j)*(+i)*(+j)=-i$. 
This process is referred to as the non-Abelian braiding of nodes.
Note that the charge $i^2=  j^2 =-1$ captures that there are two stable nodes with the same charge within one gap, while $i*(-i)=  j*(-j) =1$ suggests there are two nodes with opposite charge or no stable nodes in the gap.

Even though the frame charges, as elements of the first homotopy group, already describe the rich non-Abelian physics of the nodes, they are limited in capturing the stability of multiple pairs of nodes with equal charges in the same gap, for example, $ i^4 = j^4 = 1$ ~\cite{peng2022phonons,jiang2021experimental}. 
The problem arises because the elements of $\pi_1$ are based on the gap condition over a 1D loop. However, the information within the region enclosed by the loop is also important.
To overcome this limitation, we consider the gap condition over a 2D patch of the BZ containing the nodes to calculate the so-called patch Euler class~\cite{wu2019non,bouhon2020non,peng2022phonons}, a generalization of the Euler class defined in the whole BZ~\cite{ahn2019failure}.

{\bf Patch Euler class.} 
As we mentioned in the main text, the system with space-time inversion $I_{ST}$ symmetry $I^2_{ST}=1$ implies the reality of the Hamiltonian. 
In 2D systems, such $I_{ST}$ appears in the form of $I_{ST}=PT$ in a spinless case or $I_{ST}=C_{2z}T$ in either spinless or spinful case.
When an antiunitary operator $\mathcal{A}=U_{\mathcal{A}}\mathcal{K}$ satisfies $\mathcal{A}^2=1$, where $U_{\mathcal{A}}$ is a unitary operator and $\mathcal{K}$ is complex conjugation, we can always find a basis in which $U_{\mathcal{A}}$ becomes the identity using the Autonne-Takagi decomposition~\cite{bouhon2020non}.
Therefore, in the basis where $I_{ST}=\mathcal{K}$, $I_{ST}H({\bm{k}})I_{ST}^{-1} = H({\bm{k}})^*=H({\bm{k}})$ indicates that Hamiltonian is real at any $\bm{k}$.

In the real basis, the Euler class can be defined by the integral of Euler form over a closed surface as in Eq.~(1).
For systems with multiple pairs of nodes in the same gap, the patch Euler class~\cite{bouhon2020non,peng2022phonons} can be defined as $\chi(\mathcal{S})=\frac{1}{2 \pi}\left[\int_{\mathcal{S}} \tilde{\bm{F}}_{12}(\bm{k}) \mathrm{d} k_1 \mathrm{d} k_2-\oint_{\partial \mathcal{S}} \tilde{\bm{A}}_{12}(\bm{k}) \cdot \mathrm{d} \bm{k}\right]$ over a patch $\mathcal{S} \subset$ BZ that includes the nodes of interest.
To calculate the patch Euler class, we first do an area integration of the Euler curvature $\tilde{\bm{F}}_{12}(\bm{k})$ on a region $\mathcal{S}$ enclosing the nodes of interest, and then minus the boundary $\partial \mathcal{S}$ integration of Euler connection $\tilde{\bm{A}}_{12}(\bm{k})=\langle \tilde{u}_{1\bm{k}}|\nabla_{\bm{k}}|\tilde{u}_{2\bm{k}}\rangle$.
The nontrivial patch Euler number $\chi(\mathcal{S})$ indicates the existence of  nodes whose total charge is $2\chi(\mathcal{S})$ inside the patch $\mathcal{S}$.
Numerically, after combining the two real states to a complex state $(|\tilde{u}_{1\bm{k}}\rangle+i |\tilde{u}_{2\bm{k}}\rangle)$, the patch Euler number can be calculated similarly to how we calculate the Chern number~\cite{Fukui}.
As we can see from the definition, even if there are many nodes of two bands between the gap, we can study the patch Euler number within a small region of interest. 

\section{S2. General effective four-band Hamiltonian describing the gap closing}\label{gapclosing}

\begin{table}[h]
\caption{\textbf{Hermitian matrices allowed by the symmetries.} The $4\times4$ traceless hermitian matrices that are invariant under $M_z$ and $C_{2z}T$ symmetries are listed.}
\begin{tabular}{|c|c|c|c|}
\hline
\diagbox{$C_{2z}T$}{$M_z$} & $i\sigma_z\tau_0$ & $i\sigma_z\tau_x$ & $i\sigma_z\tau_z$ \\ \hline
$i\sigma_x\tau_0\mathcal{K}$ & $\sigma_0\tau_x$, $\sigma_0\tau_z$, $\sigma_z\tau_y$ & $\sigma_0\tau_x$, $\sigma_x\tau_z$, $\sigma_y\tau_z$ & $\sigma_0\tau_z$, $\sigma_x\tau_x$, $\sigma_y\tau_x$ \\ \hline
$i\sigma_x\tau_x\mathcal{K}$ & $\sigma_0\tau_x$, $\sigma_0\tau_y$, $\sigma_z\tau_z$ & $\sigma_0\tau_x$, $\sigma_x\tau_y$, $\sigma_y\tau_y$ & None \\ \hline
$i\sigma_x\tau_z\mathcal{K}$ & $\sigma_0\tau_y$, $\sigma_0\tau_z$, $\sigma_z\tau_x$ & None & $\sigma_0\tau_z$, $\sigma_x\tau_y$, $\sigma_y\tau_y$ \\ \hline
\end{tabular}
\label{table1}
\end{table}

    If the mirror reflection symmetry about the $xy$-plane ($M_z$) is present in addition to $C_{2z}T$, the Kramers' degeneracy is protected by $PT$ ($=M_zC_{2z}T$ and $(PT)^2=-1$) at any $k$.
    In such a case, the nodal points between the Kramers' pair bands are ill-defined, so the braiding process is hidden.
    More precisely, in the presence of $M_z$, the phase transition between the two Euler phases is not mediated by the braiding process but occurs directly between them.
    Let us consider a four-band effective Hamiltonian $H_{\textrm{eff}}(k_x,k_y)$ to describe the gap closing as follows:
    We first enlarge the parameter space of $H_{\textrm{eff}}$ by considering $p$ as a new component of the momentum.
    Then we write $H_{\textrm{eff}}(k_x,k_y,p)=f_{ij}(k_x,k_y,p)\sigma_i\tau_j$ where $\sigma_{i=0,1,2,3}$ and $\tau_{j=0,1,2,3}$ denote the Pauli matrices that represent spin and band degrees of freedom, respectively. $f_{ij}(k_x,k_y,p)$s are linear functions of $k_x$, $k_y$ and $p$.
    Within the same basis, the expression for $T$ is given by $T=i\sigma_y\mathcal{K}$.
    Since the spin part of $C_{2z}$ is $i\sigma_z$ and $(C_{2z}T)^2=1$, the possible form of $C_{2z}T$ is $i\sigma_x\tau_0\mathcal{K}$, $i\sigma_x\tau_x\mathcal{K}$, or $i\sigma_x\tau_z\mathcal{K}$.
    Given that $M_z$ commutes with $T$, $M_z$ can take one of the following forms $i\sigma_z\tau_0$, $i\sigma_z\tau_x$, and $i\sigma_z\tau_z$.
    For all possible combinations of $C_{2z}$ and $M_z$ matrix expressions, we find that $H_{\textrm{eff}}$ that is invariant under both $C_{2z}T$ and $M_z$ can be written as $H_{\textrm{eff}}(k_x,k_y,p)=a(k_x,k_y,p)\Gamma_1+b(k_x,k_y,p)\Gamma_2+c(k_x,k_y,p)\Gamma_3$ with linearly independent, mutually anticommuting Hermitian matrices $\Gamma_1$, $\Gamma_2$, and $\Gamma_3$.
    The eigenvalues of this Hamiltonian are $\pm\sqrt{a^2+b^2+c^2}$, so we have three equations to achieve a gap closing point: $a(k_x,k_y,p)=0$, $b(k_x,k_y,p)=0$, and $c(k_x,k_y,p)=0$.
    Since there are 3 questions to be satisfied in the presence of 3 variables, the gap-closing solution appears in the form of an isolated point with four-fold degeneracy.
    
    In the $(k_x,k_y,p)$ 3D space, we can define a topological charge on a 2D closed manifold enclosing the node.
    In contrast to realistic 3D systems where $C_{2z}T$ reverses the $k_z$ momentum, in this artificial 3D system, $C_{2z}T$ doesn't affect the third momentum.
    Thus, any 2D manifold in the artificial momentum space is invariant under $C_{2z}T$ so we can define a topological charge of the node by the Euler number, the integration of Euler form $\tilde{\bm{F}}(k_x,k_y,p)$ over a closed sphere enclosing the node.
    Analogous to the relation between the Weyl points and Chern numbers, the Euler number on the sphere enclosing the node at $p_c$ corresponds to the difference of Euler classes of 2D planes above ($p>p_c$) and below ($p<p_c$) the node.

    When $M_z$ is absent, the point node inflates, being three nodal rings corresponding to the gaps between the 1st-2nd, 2nd-3rd, and 3rd-4th bands.
    The cross-section of these nodal rings by a constant-$p$ plane shows the position of nodal points in the real 2D momentum space.

The sets of $\Gamma_1$, $\Gamma_2$, and $\Gamma_3$ for all possible combinations of $C_{2z}T$ and $M_z$ are summarized in Table~\ref{table1}. We note that the two combinations with no entries, $C_{2z}T=i\sigma_x\tau_x\mathcal{K}$ with $M_z=i\sigma_z\tau_z$ and $C_{2z}T=i\sigma_x\tau_z\mathcal{K}$ with $M_z=i\sigma_z\tau_x$, are forbidden physically, because $P=C_{2z}M_z$ does not commute with $T$ in such cases. This table shows that the three $\Gamma$ matrices are mutually anticommuting in every case.

\section{S3. Nodal braiding in the BHZ model}\label{BHZ}

\begin{figure*}[!h]
 \centering
 \includegraphics[width=0.95\textwidth]{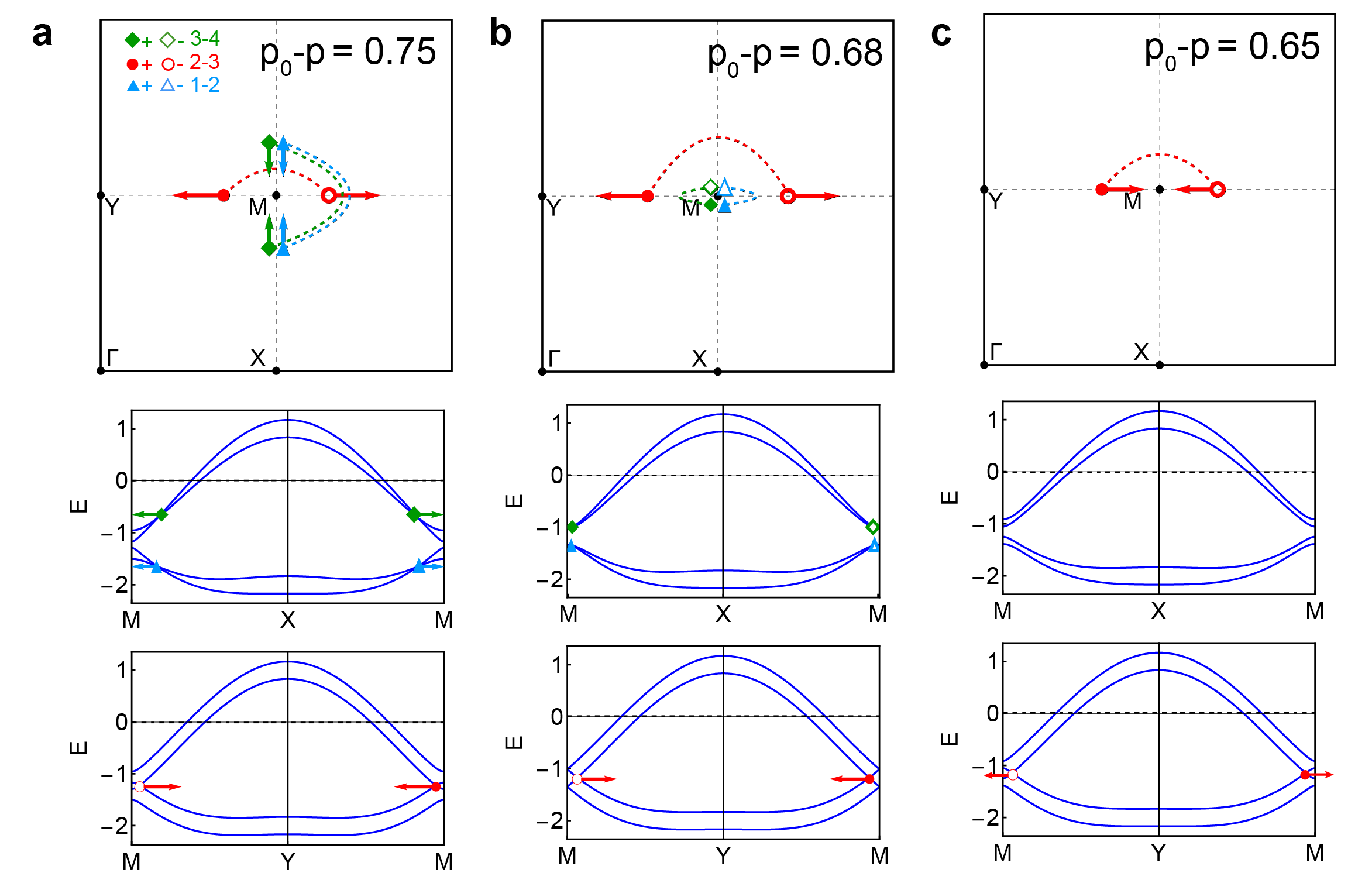}
 \caption{
        {\bf The braiding process appearing in the phase transition of the BHZ model.} In this figure, we use the parameters $t_{ss}=0.3$, $t_{sp}=0.6$, $t_{pp}=0.8$, and $m=0.2$.
        The above panels display the positions of nodes in the BZ. The following two panels illustrate the relative positions of nodes on the band structures along $k_x$ and $k_y$ directions.
        The filled (empty) blue, red, and green markers represent the nodes between the first and second, the second and third, and the third and fourth bands with vorticity $+1$ ($-1$), respectively. The dashed lines represent the Dirac strings that connect two nodes with the same color. We used $p_0-p=0.75$, 0.68, and 0.65 for \textbf{a}, \textbf{b}, and \textbf{c}, respectively.
 }
 \label{BHZfig}
 \end{figure*}

    In the main text, we mentioned that for the magnetic BHZ model, the topological phase transition from $e_2=1$ to $e_2=0$ phase is mediated by the braiding of nodes. In this section, we show the details of the nodal braiding process.
    We start from the $e_2=1$ phase. As $p$ increases, the blue (green) nodes move toward the $M$ point along the $X-M$ line.
    Meanwhile, a pair of red nodes with opposite vorticities are created at $M$ and moves away from $M$ along the $Y-M$ line (Fig.~\ref{BHZfig} (a)).
    Before the blue (green) nodes pair meets at $M$, one of them passes through the Dirac string of red nodes and changes its sign (Fig.~\ref{BHZfig} (b)).
    Consequently, when they eventually merge at $M$, blue (green) nodes open a gap due to their opposite vorticities (Fig.~\ref{BHZfig} (c)).
    During this process, one of the red nodes passes through two Dirac strings (one from the blue nodes and the other from the green nodes), thereby returning to its original vorticity.
    After the pair-annihilation of blue (green) nodes, with a further increase in $p$, the red nodes move back towards $M$, finally undergoing pair annihilation and reaching the $e_2=0$ phase.
    
\section{S4. Kagome lattice ferromagnet}\label{kagomeFM}
In this section, we study the MEI phase and its topological phase transition in the kagome lattice ferromagnet. We consider the tight-binding Hamiltonian $H_{\textrm{KFM}}(\bm k)$ that includes nearest-neighbor hoppings, spin-orbit coupling, and mean-field exchange couplings that describe in-plane ordered ferromagnetic:
\begin{equation}
    H_{\textrm{KFM}}(\bm k)=
    -t
    \begin{pmatrix}
	0 & h_{12} & h_{13} \\
	h_{21} & 0 & h_{23} \\
	h_{31} & h_{32} & 0
    \end{pmatrix}\otimes\sigma_0-i\lambda
    \begin{pmatrix}
	0 & h_{12} & -h_{13} \\
	-h_{21} & 0 & h_{23} \\
	h_{31} & -h_{32} & 0
    \end{pmatrix}
    \otimes\sigma_z-\bm{m}\cdot(I_{3}\otimes\bm\sigma),
\end{equation}
where $\bm{k}=(k_x,k_y)$, $\bm{e}_1=(1,0)$, $\bm{e}_2=(-1/2,\sqrt{3}/2)$, $\bm{e}_3=(-1/2,-\sqrt{3}/2)$, $h_{12}=(1+e^{i\bm{k}\cdot\bm{e}_1})=$, $h_{13}=(1+e^{-i\bm{k}\cdot\bm{e}_3})$, $h_{23}=(1+e^{i\bm{k}\cdot\bm{e}_2})$ and $h_{ij}=h_{ji}^*$.
In the latter calculations, we use $t=1$, $\lambda=0.1$, and $\bm{m}=(0.2,0,0)$.

\begin{figure*}[!h]
 \centering
 \includegraphics[width=0.98\textwidth]{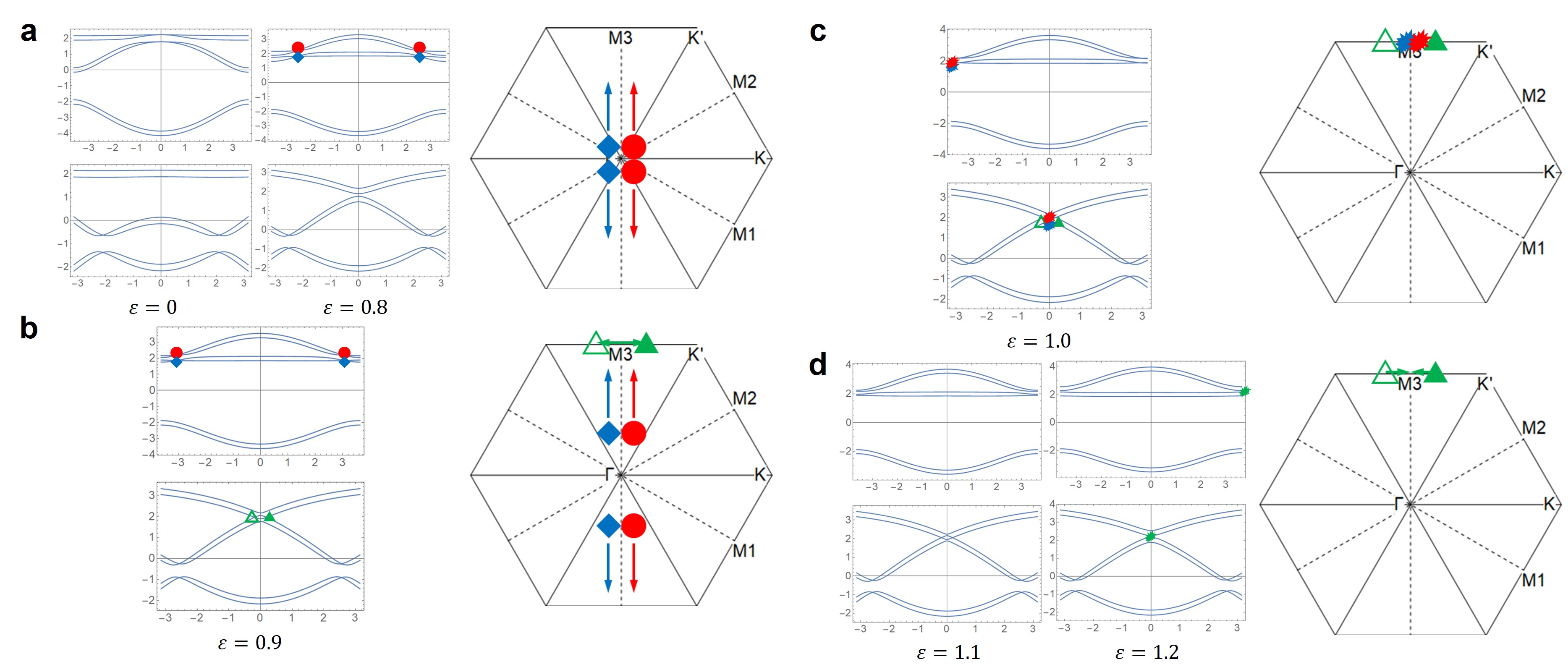}
 \caption{
        {\bf{The braiding process appears in the phase transition between $e_2=1$ ($\varepsilon=0$) and $e_2=0$ ($\varepsilon=1.2$) phases.}} The filled (empty) blue, green, and red markers represent the nodes between the third and fourth, the fourth and fifth, and the fifth and sixth bands with vorticity $+1$ ($-1$), respectively.
        From \textbf{a} to \textbf{d}, the upper band structures show the dispersion along the $M_3-\Gamma-M_3$ line while the lower band structures show the dispersion along the $K-M_3-K'$ line.
        The sublattice potential used for each figure is shown below the band structures.
 }
 \label{KFMfig}
 \end{figure*}

As shown in Fig.~\ref{KFMfig} (a), the two nearly flat bands at the highest energies have $e_2=1$, while the two lowest bands have $e_2=1$ as well. Consequently, a quadratic band touching between the highest two bands with vorticity 2 appears at the $\Gamma$ point, while linear band crossings between the lowest two bands with vorticity 1 appear at the $K$ and $K'$ points.
We then investigate the braiding process during the transition from the MEI phase to a trivial one.
One way to achieve trivial flat bands is to introduce a large sublattice potential to one of the sublattices.
For instance, let us consider the sublattice potential term in the form of
\begin{equation}
    \begin{pmatrix}
	0 & 0 & 0 & 0 & 0 & 0 \\
	0 & 0 & 0 & 0 & 0 & 0 \\
	0 & 0 & \varepsilon & 0 & 0 & 0 \\
	0 & 0 & 0 & 0 & 0 & 0 \\
	0 & 0 & 0 & 0 & 0 & 0 \\
	0 & 0 & 0 & 0 & 0 & \varepsilon
    \end{pmatrix}.
\end{equation}
In the strong sublattice potential limit ($\varepsilon\rightarrow\infty$), the third sublattice becomes decoupled from the others. In such a case, the system can be seen as a superposition of two independent lattices: one made up of the third sublattice and the other made up of the first and second sublattices. The electrons at the third sublattice form a two-band atomic insulator with flat dispersion separated from the other bands by a large gap. These flat bands are by definition topologically trivial~\cite{bradlyn2017topological}. Therefore, by increasing $\varepsilon$ from zero to a large value, we can describe the intermediate steps of the flat bands' phase transition.

As $\varepsilon$ increases, the nodes with quadratic dispersions at $\Gamma$ split into pairs of linear band crossings (the red and blue nodes in Fig.~\ref{KFMfig} (a)) with the same vorticities and move toward $M_3$. Before they reunite at $M_3$, a pair of green nodes are created in the adjacent gap. One red (blue) node moves across the Dirac string of the green nodes and then meets the other red (blue) node at $M_3$. After the red and blue nodes are pair-annihilated, the green nodes merge at $M_3$ point and open a gap arriving at the $e_2=0$ phase.

\section{S5. Kagome lattice non-collinear antiferromagnet}\label{kagomeAFM}
\begin{figure*}[!h]
 \centering
 \includegraphics[width=0.49\textwidth]{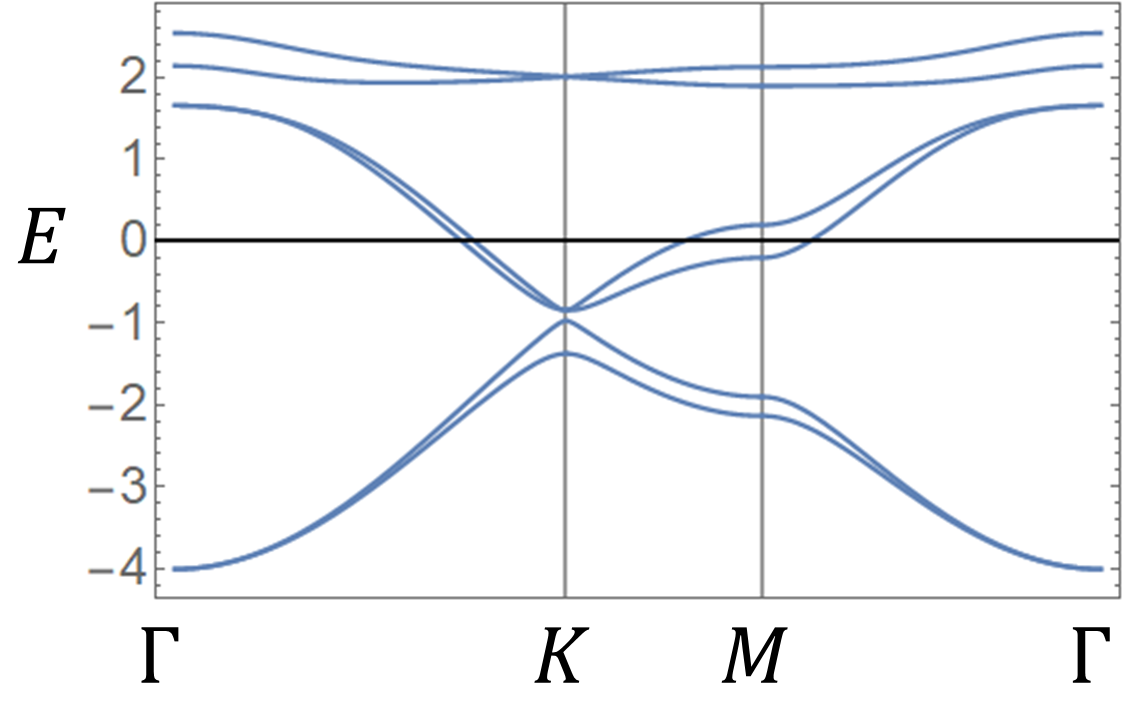}
 \caption{
     {\bf The band structure of $H_{\textrm{KAFM}}(\bm{k})$}.
 }
 \label{KAFMfig}
 \end{figure*}

The classical ground state of the antiferromagnetic Heisenberg model on the kagome lattice with Dzyaloshinskii-Moriya interaction is coplanar 120$^\circ$ ordered states. In these coplanar-but-non-collinear kagome antiferromagnets, $C_{2z}$ does not interchange the sublattices and local spin moments are all lying in the 2D plane. Therefore, they are always invariant under $C_{2z}T$. Let us consider the Hamiltonian
\begin{align}
    H_{\textrm{KAFM}}(\bm k)&=
    -t
    \begin{pmatrix}
	0 & h_{12} & h_{13} \\
	h_{21} & 0 & h_{23} \\
	h_{31} & h_{32} & 0
    \end{pmatrix}\otimes\sigma_0-i\lambda
    \begin{pmatrix}
	0 & h_{12} & -h_{13} \\
	-h_{21} & 0 & h_{23} \\
	h_{31} & -h_{32} & 0
    \end{pmatrix}
    \otimes\sigma_z-\\
    &\begin{pmatrix}
	0 & 0 & 0 & m_{1x}-im_{1y} & 0 & 0 \\
	0 & 0 & 0 & 0 & m_{2x}-im_{2y} & 0 \\
	0 & 0 & 0 & 0 & 0 & m_{3x}-im_{3y} \\
	m_{1x}+im_{1y} & 0 & 0 & 0 & 0 & 0 \\
	0 & m_{2x}+im_{2y} & 0 & 0 & 0 & 0 \\
	0 & 0 & m_{3x}+im_{3y} & 0 & 0 & 0
    \end{pmatrix},
\end{align}
where $m_{1x}=0.2\cos{7\pi/6}$, $m_{1y}=0.2\sin{7\pi/6}$, $m_{2x}=0.2\cos{-\pi/6}$, $m_{2y}=0.2\sin{-\pi/6}$, $m_{3x}=0.2\cos{\pi/2}$, and $m_{3y}=0.2\sin{\pi/2}$.
Similar to the ferromagnetic case, the lowest two bands and highest two bands of $H_{\textrm{KAFM}}(\bm k)$ have $e_2=1$. However, the nodes appear differently. In the antiferromagnetic case, as shown in Fig.~\ref{KAFMfig} the highest two bands have two nodes at $K$ and $K'$ with vorticity 1 for each, while the lowest two bands have a single node with vorticity 2 at $\Gamma$.

\section{S6. Topology of ZrTe$_5$ systems}\label{app:zrteband}

 \begin{figure*}[!h]
 \centering
 \includegraphics[width=0.7\textwidth]{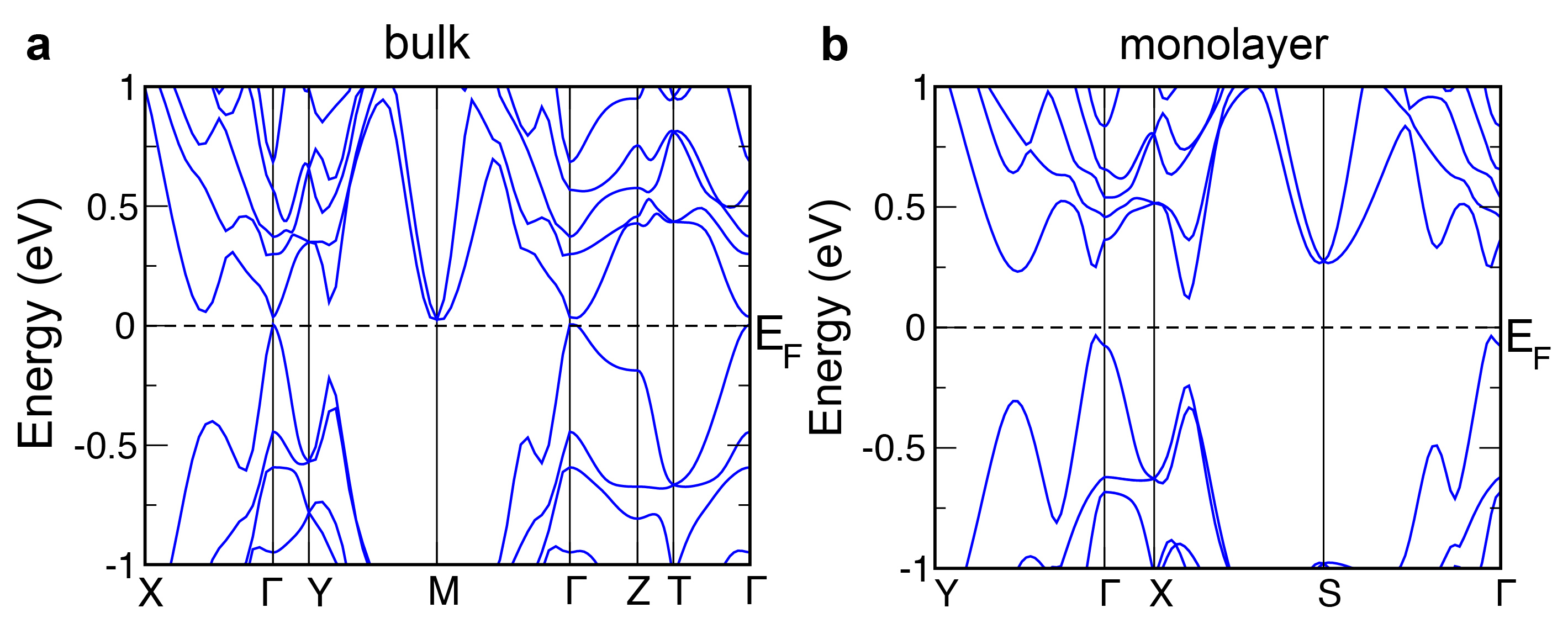}
 \caption{
 {\bf Band structure of {\bf a.} bulk and {\bf b.} monolayer ZrTe$_5$.}
 }
 \label{figs:bulk}
 \end{figure*}

In this section, we first reproduce the electronic property of nonmagnetic bulk and monolayer ZrTe$_5$ and then analyze the topology of bilayer ZrTe$_5$.
As shown in Fig.~\ref{figs:bulk}, the band structures of bulk and single layer ZrTe$_5$ are well consistent with the previously reported results~\cite{weng2014}.
For bulk ZrTe$_5$ with space group $Cmcm$ (No.63), we first check the band structure for experimental structure~\cite{zrte_str}. It shows the system is a strong topological insulator (TI) with ($v_1,v_2,v_3;v_0$)=(110;1).
After relaxation, the crystal structure becomes $1.4\%$ larger than the experimental ones, and the system becomes a 3D weak TI with ($v_1,v_2,v_3;v_0$)=(110;0).
For monolayer ZrTe$_5$ with space group $Pmmn$ (No.59), the system is a quantum spin Hall insulator with $\nu_2=1$. 
%
%
Since the lattice parameters and atomic position are very close to the original phase and the topology of monolayer ZrTe$_5$ doesn't change after relaxation, we construct bilayer ZrTe$_5$ based on the experimental structure.

 \begin{figure*}[h!]
 \centering
 \includegraphics[width=0.98\textwidth]{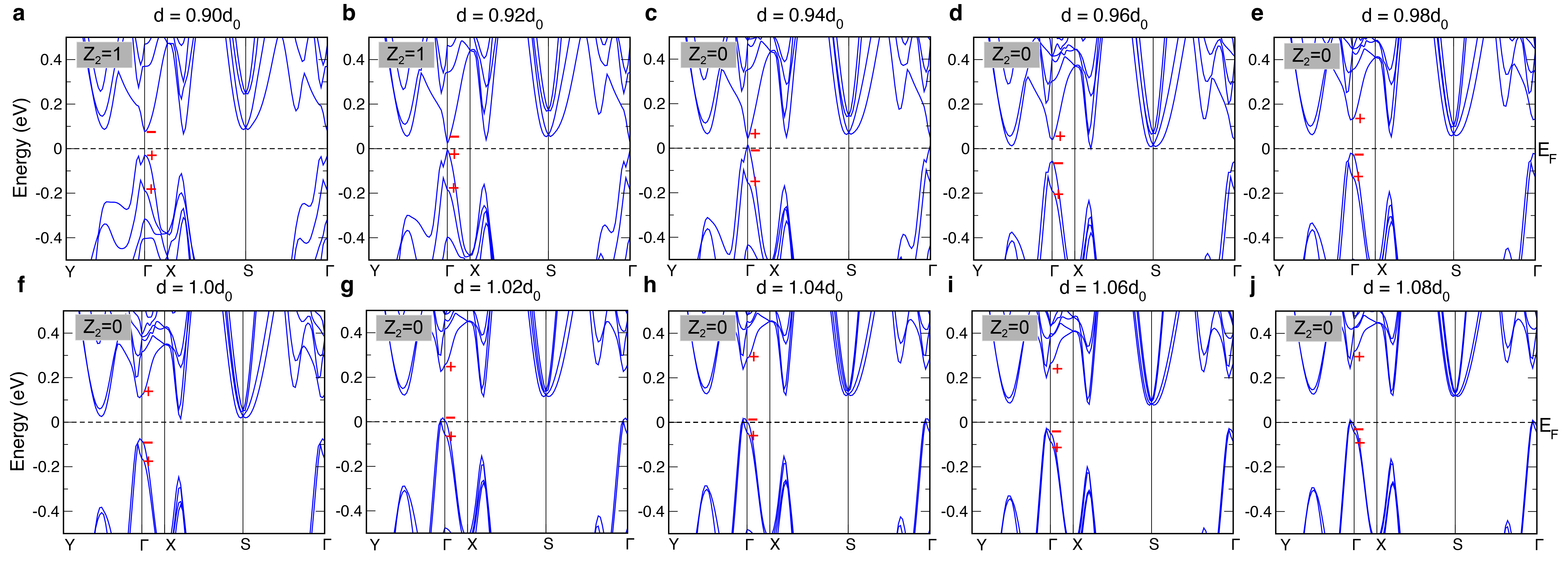}
 \caption{
 {\bf Band structure of Nonmagnetic Bilayer ZrTe$_5$ with tuning layer distance.}
 Inversion eigenvalues at $\Gamma$ are marked by the "$+/-$" sign. Band inversion occurs in between $d=0.92d_0$ and $d=0.94d_0$.
 }
 \label{figs:bilayer}
 \end{figure*}

Next, we will check the property of the nonmagnetic bilayer ZrTe$_5$.
The symmetry group of the nonmagnetic phase is $Pmma$ (No.51) with inversion symmetry.
We show the band structure of the nonmagnetic bilayer ZrTe$_5$ phase in Fig.~\ref{figs:bilayer}.
Each band is two-fold degenerated due to the combination of time-reversal symmetry and inversion symmetry.
We mark the inversion eigenvalues of the three two-fold degenerated bands near the Fermi level by "+/-" sign.
As we have shown above, monolayer ZrTe$_5$ has $\nu_2=1$.
Therefore, if two monolayers ZrTe$_5$ are stacked along the $z$ direction with weak coupling, the system will be trivial since the new system is just double of the original single-layer system (Fig.~\ref{figs:bilayer} (f)).
However, to increase the interlayer coupling by decreasing the layer distance, the system can be tuned to a topological nontrivial quantum spin Hall insulator.
As we can see, when we tune the layer distance of ZrTe$_5$, there is a band inversion in between $d=0.92d_0$ and $d=0.94d_0$, which makes a phase transition in ZrTe$_5$.
$\nu_2=1$ ($\nu_2=0$) for $d\leq 0.92d_0$ ($d\geq 0.94d_0$) can be determined by inversion eigenvalues.
However, in this nonmagnetic case, the non-Abelian braiding during the phase transition process is hidden due to the Kramers' degeneracy as we mentioned in the main text.

 \begin{figure*}[h!]
 \centering
 \includegraphics[width=0.8\textwidth]{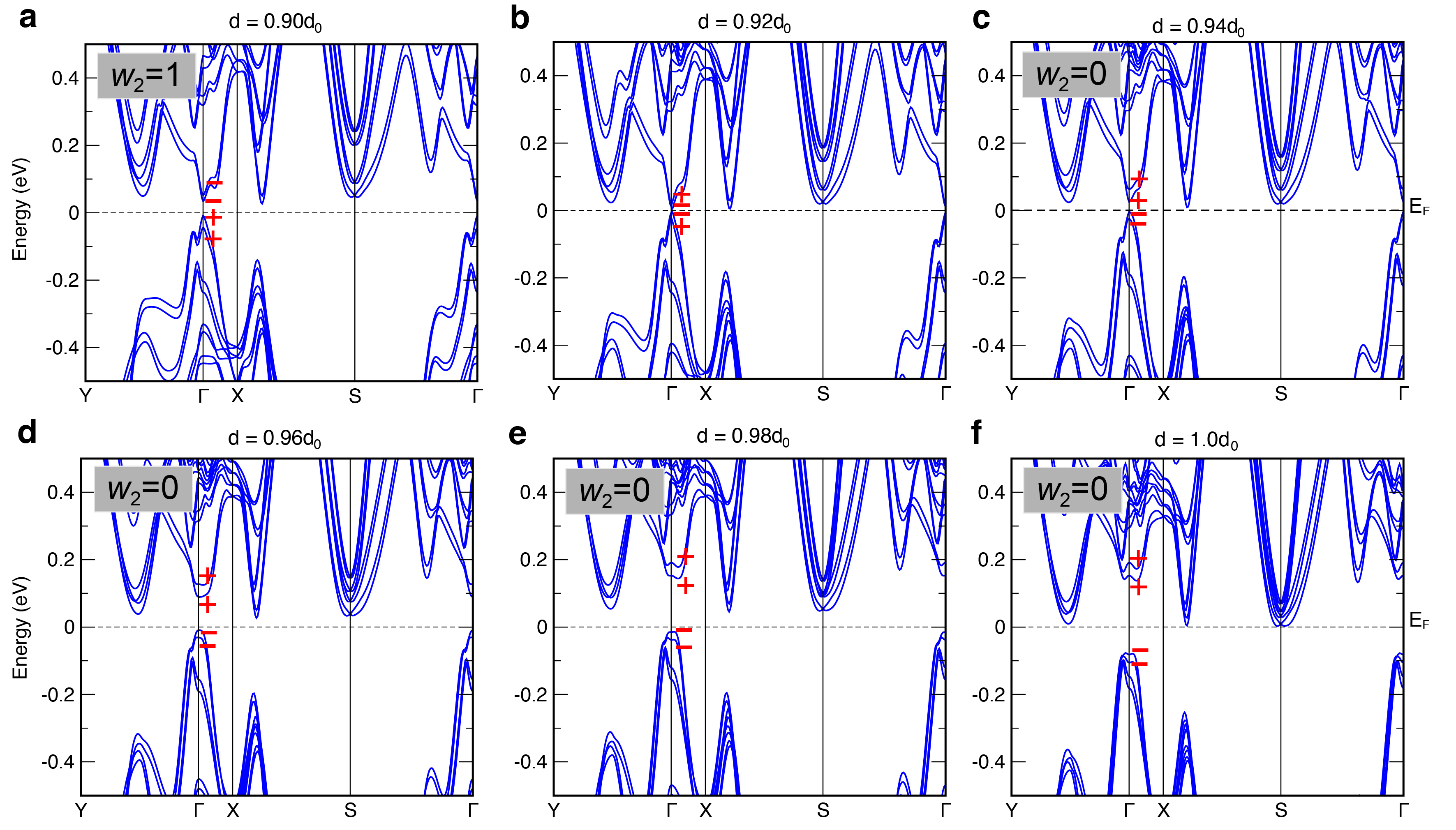}
 \caption{
 {\bf Band structure of Bilayer ZrTe$_5$ under an external magnetic field with tuning layer distance.}
 Inversion eigenvalues at $\Gamma$ are marked by the "$+/-$" sign. Band inversion occurs in between $d=0.90d_0$ and $d=0.94d_0$.
 }
 \label{figs:bilayer-mag}
 \end{figure*}

After applying an in-plane external magnetic field along (110)-direction on the bilayer ZrTe$_5$, the system is described by magnetic space group $P2'/c'$ (No.13.69). 
There are inversion $P=\{P|0,0,0\}$, \ct~=
$\{C_{2z}|0.5,0,0\}T$, and $\tilde M_zT=\{M_{z}|0.5,0,0\}T$.
Due to the absence of $PT$-symmetry, the double degeneracy at each $k$-point is lifted.
As shown in Fig.~\ref{figs:bilayer-mag}, a similar phase transition can be observed between $d=0.90d_0$ and $d=0.94d_0$. 
However, due to the breaking of time-reversal symmetry $T$ and preserving of \ct~symmetry, the system can be described by the SW class. 
By checking the inversion eigenvalues and Wilsonloop spectrum, we confirm that the system undergoes a phase transition from $w_2=0$ to $w_2=1$ through a gapless region ($d=0.90d_0 \sim 0.94d_0$).
As we already discussed in the main text, there must be a braiding process in the gapless region.
On the other hand, due to the lifting of double degeneracy at each $k$-point, the braiding process can be observed as shown in Fig.~\ref{fig:braid_mat} of the main text.

\section{S7. Magnetic Euler band topology in HgF$_2$ systems}\label{app:hgf2band}

 \begin{figure*}[t!]
 \centering
 \includegraphics[width=0.98\textwidth]{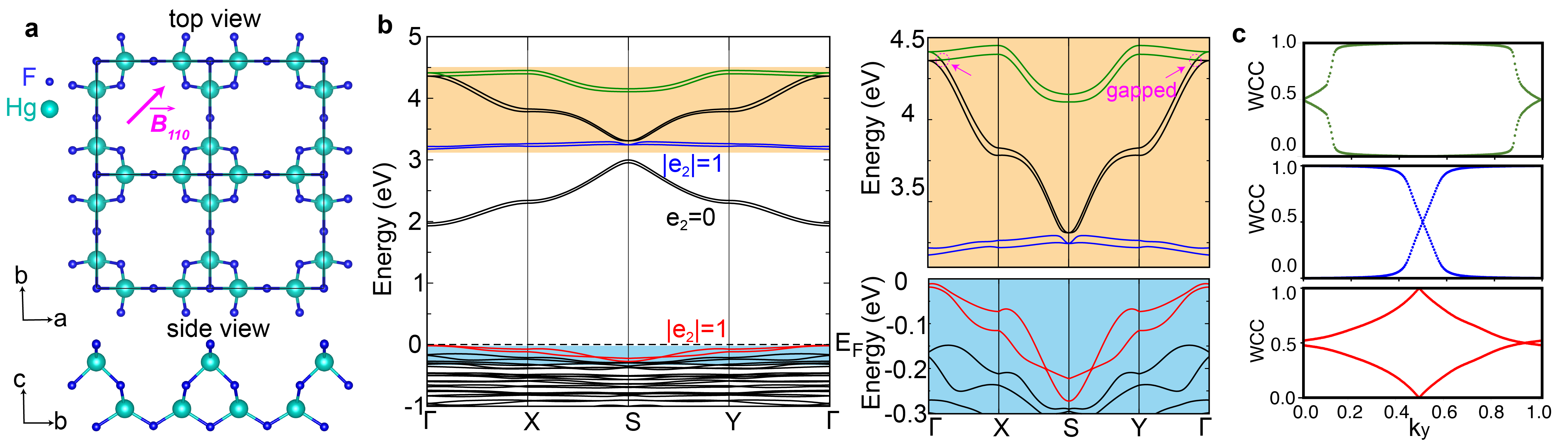}
 \caption{
 {\bf HgF$_2$ with Euler topology under magnetic field.}
 (a) Crystal structure of diamond-octagon lattice HgF$_2$ considering in-plane magnetism along (110)-direction.
 (b) Band structure of HgF$_2$ with spin-orbit coupling. Close-ups of the blue and orange regions in the left panel are shown in the right panel.
 (c) $k_y$-directed Wilsonloop spectra for each of the two red, blue, and green bands in (b).
 }
 \label{fig:hgf}
 \end{figure*}

Nonmagnetic monolayer $M$F$_2$ family ($M$=Zn, Cd and Hg) has been unveiled to exhibit a unique diamond-octagon lattice structure (i.e., line-graph lattice of Lieb lattice) with space group $P$-4$m$2~\cite{2dmatpedia,HgF2prm} as shown in Fig.~\ref{fig:hgf} (a).
Concurrently, recent studies have identified the presence of dual topological flat bands within these systems~\cite{HgF2prm}.
According to our theory, by introducing an in-plane Zeeman field, it becomes possible to realize the Euler topology in these materials. To illustrate this concept, let's consider a specific material \hgf.

Before and after applying the external magnetic field, the band structure of the system changes slightly.
After considering an in-plane magnetic field along (110)-direction, the band structure of \hgf ~with spin-orbit coupling (SOC) is obtained from DFT shown in Fig.~\ref{fig:hgf} (b).
Evidently, eight bands situated above the Fermi energy align with the diamond-octagon lattice structure. Notably, dual flat bands are represented by the colors blue and green.
Furthermore, two isolated bands are located near the Fermi level marked by red.
We note that before applying the external magnetic field, even though the total occupied bands are topologically trivial, the dual flat bands and two red high-occupied bands in the diamond-octagon lattice are $\mathbb{Z}_2$ nontrivial \cite{palprb}.
This indicates that if the chemical potential is just above blue bands (around 3.25 eV), the system can be considered as a quantum spin Hall insulator.
To identify their topological properties under the external magnetic field, we calculated the Wilson loop spectra.
Since the in-plane magnetic field breaks both the $C_{2z}$ symmetry and the time-reversal symmetry $T$, while preserving the \ct~symmetry, the system can be described by a real Hamiltonian.
The topology of any isolated two bands for this real Hamiltonian is characterized by the Euler class.
Remarkably, the results in Fig.~\ref{fig:hgf}(c) reveal a nontrivial Euler class with $|e_2|=1$ for each pair of bands implying the existence of two band crossings with the same winding which cannot be annihilated.
According to our theory, if the chemical potential is chosen around 3.25 eV, that quantum spin Hall insulator turns into a magnetic Euler insulator under in-plane magnetism.
%
%
We note that even if chemical potential is chosen at the Fermi level, this system is also a magnetic Euler insulator because the two isolated $\mathbb{Z}_2$ nontrivial red bands can turn into two Euler bands under in-plane magnetism. 
In the main text, we propose that one can always obtain a magnetic Euler/SW insulator after considering an in-plane magnetic field on QSHI.  However, this exemplification further illustrates that even starting from a trivial insulator, it is possible to observe the emergence of a magnetic Euler insulator when considering in-plane magnetism, provided that the two highest occupied bands are $\mathbb{Z}_2$ nontrivial.
This system provides a stable platform to study the Euler topology further in experiments.



\end{document}